\documentclass[a4paper,fleqn]{cas-sc}
\usepackage[numbers,sort&compress]{natbib}
\usepackage{setspace}
\usepackage{threeparttable}
\usepackage{hyperref}
\begin{document}
\let\WriteBookmarks\relax
\def\floatpagepagefraction{1}
\def\textpagefraction{.001}

\shortauthors{Lei et al.}  
\title [mode = title]{GPINND: A deep-learning-based state of health estimation for Lithium-ion battery}

\address{College of Information Science and Electronic Engineering, Zhejiang University, Hangzhou 310027, China}

\author{Yuzhu Lei}[orcid=0009-0001-5268-1915]
\cormark[1]
\cortext[mycorrespondingauthor]{Corresponding author}
\ead{leiyuzhu@zju.edu.cn}

\author{Guanding Yu}[orcid=0000-0001-7296-1490]

\begin{abstract}
Electrochemical models offer superior interpretability and reliability for battery degradation diagnosis. However, the high computational cost of iterative parameter identification severely hinders the practical implementation of electrochemically informed state of health (SOH) estimation in real-time systems. To address this challenge, this paper proposes an SOH estimation method that integrates deep learning with electrochemical mechanisms and adopts a sequential training strategy. First, we construct a hybrid-driven surrogate model to learn internal electrochemical dynamics by fusing high-fidelity simulation data with physical constraints. This model subsequently serves as an accurate and differentiable physical kernel for voltage reconstruction. Then, we develop a self-supervised framework to train a parameter identification network by minimizing the voltage reconstruction error. The resulting model enables the non-iterative identification of aging parameters from external measurements. Finally, utilizing the identified parameters as physicochemical health indicators, we establish a high-precision SOH estimation network that leverages data-driven residual correction to compensate for identification deviations. Crucially, a sequential training strategy is applied across these modules to effectively mitigate convergence issues and improve the accuracy of each module. Experimental results demonstrate that the proposed method achieves an average voltage reconstruction root mean square error (RMSE) of 0.0198 V and an SOH estimation RMSE of 0.0014. 
\end{abstract}

\begin{highlights}
\item A sequential training strategy is proposed to decouple physically coupled heterogeneous tasks.
\item A hybrid-driven surrogate model serves as an accurate and differentiable physical kernel.
\item An electrochemically informed network enables online aging parameter identification without iteration.
\item Aging parameters are leveraged as physicochemical HIs to improve SOH interpretability and accuracy.
\item GPINND achieves an SOH estimation RMSE of 0.0014 with millisecond-level inference speed.
\end{highlights}


\maketitle
\section{Introduction} \label{Section 1}
Lithium-ion batteries have been widely used in grid energy storage, electric vehicles, and portable electronic products due to their high energy density and long cycle life~\cite{r37,r38,r39}. However, complex physicochemical changes inevitably occur during battery cycling, including loss of active material (LAM) and loss of lithium-ion inventory (LLI)~\cite{r40,r86,r43}. These degradation mechanisms lead to multi-scale variations. Microscopically, key aging parameters, including the active material volume fractions and the lithium inventory, progressively deteriorate~\cite{r41,r42}. Macroscopically, the available capacity suffers continuous decay~\cite{r59,r97}. Typically, the state of health (SOH) is defined as the ratio of current available capacity to initial capacity, reflecting the degree of battery performance degradation. Accurate SOH assessment not only facilitates intelligent energy management to extend battery lifespan, but also provides the essential basis for sorting retired batteries for effective cascaded utilization~\cite{r85,r82,r89}. Therefore, developing efficient, robust, and precise SOH estimation methods is paramount for battery management systems (BMS)~\cite{r87}.

Existing SOH estimation methods can be mainly divided into model-driven methods, data-driven methods, and model-data fusion methods~\cite{r68,r81}. Model-driven approaches simulate battery dynamics by mathematical models and utilize optimization algorithms to identify model parameters~\cite{r66}. Common models include electrochemical models, equivalent circuit models (ECMs), and empirical models. Based on first principles, the electrochemical models describe the internal battery dynamics and underlying degradation mechanisms~\cite{r83}. For instance, Kim et al. employed the genetic algorithm to identify parameters of the pseudo-two-dimensional (P2D) model for degradation diagnosis~\cite{r75}. The resulting root mean square error (RMSE) of reconstructed voltages with the identified parameters is 18.79 mV. However, the iterative parameter optimization process is computationally intensive, as it requires repeated numerical solutions of the electrochemical model. Fan et al. used the particle swarm optimization algorithm to identify parameters of the single particle model with electrolyte (SPMe), reducing the voltage reconstruction RMSE to 11 mV~\cite{r36}. Notably, a single identification required 30 minutes to converge, severely limiting the application of such methods in real-time BMS. Similarly, Li et al. employed the cuckoo search for parameter identification of the P2D model~\cite{r96}. This approach required approximately 15 hours to identify 26 parameters. In contrast, while ECMs and empirical models offer higher computational efficiency, they fail to provide insights into microscopic degradation mechanisms.

In recent years, breakthroughs in deep learning have driven the rapid development of data-driven SOH estimation methods~\cite{r35,r88}. Generally, these methods estimate SOH by mining the mapping between operating data (e.g., voltage, current, and temperature) and battery capacity~\cite{r58,r90}. For example, Bai et al. employed convolutional neural networks and Transformer to extract features and estimated SOH via a multi-head attention module~\cite{r54}. Wei et al. extracted diverse health indicators (HIs) through differential thermo-voltammetry and incremental capacity analysis~\cite{r47}. They then employed the principal component analysis for feature selection and ultimately fed highly correlated features into a group data processing neural network to realize accurate SOH estimation. However, while purely data-driven models perform well in fitting accuracy, the inherent black-box nature of neural networks (NNs) leads to a lack of mechanistic interpretability, which severely limits the reliability of data-driven approaches in practical applications~\cite{r91,r92}.

Model-data fusion methods facilitate the enhancement of comprehensive performance by synergizing the strengths of model-driven and data-driven approaches~\cite{r56,r57,r67}. A common approach is to use the identified parameters of ECMs as inputs for neural networks~\cite{r50,r51,r52}. For example, Chen et al. fed parameters such as internal resistance into NNs for SOH estimation to improve model robustness~\cite{r50}. Although this approach offers better interpretability than pure data-driven models, ECM parameters do not represent the underlying microscopic degradation mechanisms. 
Furthermore, some researchers construct physics-informed neural networks (PINNs) by incorporating capacity degradation empirical formulas as physical constraints to improve SOH estimation performance. For example, Liu et al. employed a semi-empirical equation of capacity decay caused by solid electrolyte interface (SEI) growth to guide SOH estimation model training~\cite{r61}. Wang et al. proposed using an auxiliary NN to model the governing differential equation of SOH evolution, which serves as a physical constraint for model training~\cite{r65}. Wen et al. proposed two PINN architectures utilizing the Verhulst formula and the deep hidden physics model (DeepHPM), respectively, as physical losses~\cite{r63}. These methods enhance model generalization capability through physical constraints~\cite{r77}. However, as these studies mostly rely on empirical equations or macroscopic evolution patterns, they still exhibit limited capability to elucidate the underlying degradation mechanisms. Ideally, microscopic parameters of electrochemical models are robust and  mechanistically interpretable HIs for SOH estimation. Nevertheless, their online identification is blocked by the time-consuming iterative parameter optimization.

Recently, some studies have attempted to integrate electrochemical models into deep learning frameworks, aiming to achieve computationally efficient and mechanistically interpretable SOH estimation. Essentially, these frameworks embed physical governing equations directly into NNs as training constraints, allowing the networks to learn internal electrochemical dynamics. Under such constraints, these studies employ multi-task learning to simultaneously perform internal concentration dynamics prediction, parameter identification, and SOH estimation. Crucially, the resulting parameter identification network, trained with electrochemical constraints, enables direct, non-iterative parameter identification. Consequently, this enables efficient online microscopic analysis and physics-informed SOH estimation. For instance, Zhang et al. constructed a physics-informed network constrained by the complex partial differential equations (PDEs) of the SPMe model. This comprehensive framework enables the simultaneous estimation of lithium-ion concentration, electrode aging parameters, and capacity degradation~\cite{r60}. However, directly embedding complex PDEs often leads to vanishing or exploding gradients during network training. To mitigate this complexity, simplified ordinary differential equations (ODEs) governing solid-liquid diffusion were employed as constraints in~\cite{r55}. Combined with voltage reconstruction and capacity estimation errors derived from the predicted lithium-ion concentration dynamics for multi-task training, this approach enables electrochemical aging-informed SOH estimation. Similarly, simplified constraints were applied to predict internal concentration dynamics in~\cite{r71}. Subsequently, incremental capacity analysis and differential voltage analysis were utilized to extract electrode-level aging features for battery health prediction. Despite the advancements, in these studies, the joint optimization of the physically coupled heterogeneous tasks often suffers from gradient conflicts and convergence issues, which leads to limited accuracy~\cite{r95}. Besides, relying on simplified ODEs as physical constraints inevitably introduces systematic bias in internal concentration prediction and leads to cumulative errors in voltage reconstruction and SOH estimation. This misguides the parameter optimization process and ultimately compromises the SOH estimation performance. Consequently, the key challenge limiting performance breakthroughs lies in how to effectively improve model precision while ensuring the collaborative convergence across these tasks. 

To this end, we propose an SOH estimation method that integrates deep learning with electrochemical mechanisms and adopts a sequential training strategy, referred to as GPINND. The heterogeneous tasks, including internal dynamics prediction, parameter identification, and SOH estimation, inherently follow a physical progression. Therefore, a sequential training strategy is a logical and effective solution~\cite{r93}. This approach allows for the sophisticated and task-specific design of each module to enhance its individual performance, while ensuring robust convergence at each step. It directly addresses the core challenge of improving model accuracy and maintaining collaborative stability across the entire framework. Following this sequential training strategy,  our proposed framework is structured into three modules. First, we construct a hybrid-driven surrogate model to learn internal concentration dynamics, which provides the essential electrochemical basis for training the subsequent online parameter identification network. Based on this surrogate model, a parameter identification network is trained by minimizing the voltage reconstruction error. Finally, leveraging the identified aging parameters as physicochemical HIs, we establish a high-precision SOH estimation network. The main contributions of this paper are summarized as follows.

(1) We propose a sequential training strategy to decouple the physically coupled tasks, which effectively mitigates convergence issues and facilitates the specialized optimization of individual modules.

(2) We develop a hybrid-driven surrogate model that fuses high-fidelity simulation data with ODE constraints to overcome the systematic bias of simplified constraints.

(3) We construct an online parameter identification network that enables direct and non-iterative estimation of aging parameters by minimizing the voltage reconstruction error.

(4) We establish an SOH estimation network to map the identified parameters directly to battery SOH, leveraging data-driven residual correction to compensate for identification deviations.

(5) The proposed method achieves robust parameter identification with a low voltage RMSE of 0.0198 V and high-accuracy SOH estimation with an RMSE of 0.0014, demonstrating the effectiveness of the proposed framework.

The remainder of this paper is organized as follows. Section~\ref{Section 2} presents the electrochemical model and problem formulation. Section~\ref{Section 3} details the proposed GPINND framework. Section~\ref{Section 4} describes the experimental settings. Section~\ref{Section 5} provides a comprehensive discussion on the experimental results and performance validation. Finally, conclusions are summarized in Section~\ref{Section 6}.

\section{Electrochemical model and problem formulation} \label{Section 2}
\subsection{Electrochemical model}
To ensure modeling accuracy and control computational complexity, we employ the SPMe model to characterize the internal battery dynamics. The SPMe model incorporates dynamic descriptions of electrolyte lithium-ion concentration and potential, thereby overcoming the accuracy limitations of the SPM model under high-rate conditions.

\begin{figure}[htbp]
    \centering
    \includegraphics[width=.5\linewidth]{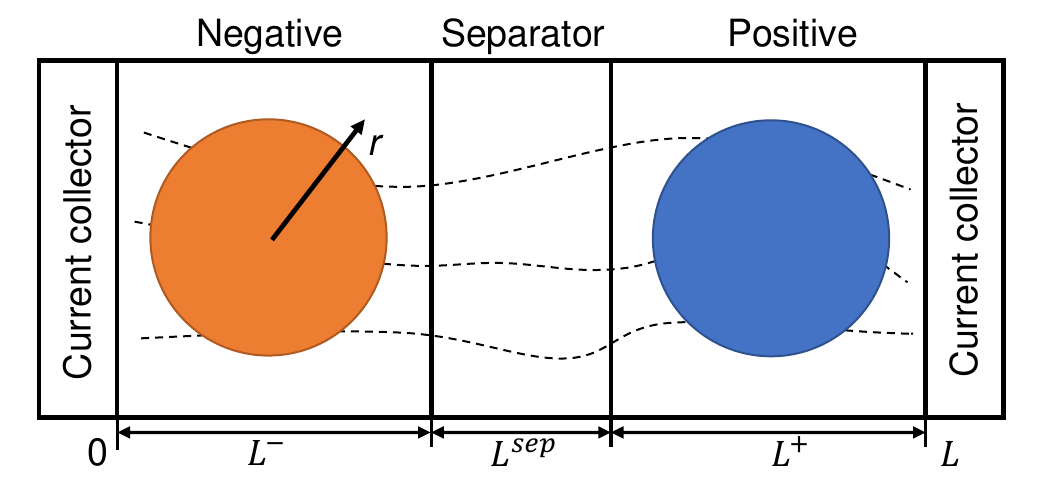}
    \caption{\label{fig:1} Schematic diagram of the SPMe model.}
\end{figure}

As shown in Fig.~\ref{fig:1}, the electrodes are approximated as single spherical active particles in the SPMe model. According to Fick's law, the solid-phase diffusion of lithium ions can be expressed as
\begin{align}
&\frac{\partial{c_s^{\pm}}\left ({r,t}\right )}{\partial t}=\frac{{D_s^{\pm}}}{r^2}\frac{\partial}{\partial r}\left ({r^2\frac{\partial{c_s^{\pm}}\left ({r,t}\right )}{\partial r}}\right ), \label{eq:1} \\
&{{\left .{\frac{\partial c_s^\pm \left ({r,t}\right )}{\partial r}}\right |}_{r=0}}=0, \label{eq:2} \\ 
&D_s^\pm{{\left .{\frac{\partial c_s^\pm \left ({r,t}\right )}{\partial r}}\right |}_{r={R_s^{\pm}}}}=-\frac{{j^{\pm}}\left ({t}\right )}{{a_s^{\pm}}F}, \label{eq:3}
\end{align}
where ${c_s^{\pm}}\left ({r,t}\right )$ denotes the solid-phase lithium-ion concentration, ${D_s^{\pm}}$ represents the solid-phase diffusion coefficient, $R_s^{\pm}$ is the particle radius, ${j^{\pm}}\left ({t}\right )$ is the volumetric reaction current density, $F$ is the Faraday constant, and ${a_s^{\pm}}$ denotes the specific interfacial area. For spherical particles, ${a_s^{\pm}}=3\varepsilon_s^{\pm}/R_s^{\pm}$, where $\varepsilon_s^{\pm}$ is the volume fraction of active material.

Commonly, the volumetric reaction current density is assumed to be uniform within the electrodes
\begin{equation}
{j^{\pm}}\left ({t}\right )=\pm \frac{I\left ({t}\right )}{{A}{L^{\pm}}}, \label{eq:4}
\end{equation}
where $I(t)$ denotes the input current, $A$ represents the electrode area, and $L^{\pm}$ is the electrode thickness.

Based on the Butler-Volmer equation, the overpotential $\eta^{\pm}(x,t)$ can be formulated as
\begin{equation}
{\eta^{\pm}}\left ({x,t}\right )=\frac{2RT}{F}\mathrm{arsinh}\left [{\frac{{j^{\pm}}\left ({t}\right )}{2{a_s^{\pm}}{i_0^{\pm}}\left ({x,t}\right )}}\right ], \label{eq:5}
\end{equation}
where $R$ and $T$ are the gas constant and temperature, respectively. Additionally, ${i_0^{\pm}}\left ({x,t}\right )$ represents the exchange current density, which can be defined as
\begin{equation}
{i_0^{\pm}}\left ({x,t}\right )=F{k_0^{\pm}}\sqrt{{{{c_e^{\pm}}\left ({x,t}\right )}}{{\left [{{c_{s,max}^{\pm}}-{c_{ss}^{\pm}}\left ({t}\right )}\right ]}}{{{c_{ss}^{\pm}}\left ({t}\right )}}}, \label{eq:6}
\end{equation}
where $k_0^{\pm}$ represents the reaction rate constant, ${{c_e^{\pm}}\left ({x,t}\right )}$ denotes the liquid-phase lithium-ion concentration, ${c_{s,max}^{\pm}}$ is the maximum solid-phase lithium concentration, and ${c_{ss}^{\pm}}\left ({t}\right )$ represents the particle surface concentration.

The liquid-phase lithium-ion concentration dynamics are described by the concentrated solution theory
\begin{align}
&\frac{\partial{c_e^{\pm}}\left ({x,t}\right )}{\partial t}=\frac{\partial}{\partial x}\left [{\frac{{D_{e,\text{eff}}^{\pm}}}{{\varepsilon_e^{\pm}}}\frac{\partial{c_e^{\pm}}\left ({x,t}\right )}{\partial x}}\right ]+ \frac{1-{t_c^{0}}}{{\varepsilon_e^{\pm}}F}{j^{\pm}}\left ({t}\right ), \label{eq:7}  \\ 
&\frac{\partial{c_e^{sep}}\left ({x,t}\right )}{\partial t}=\frac{\partial}{\partial x}\left [{\frac{{D_{e,\text{eff}}^{sep}}}{{\varepsilon_e^{sep}}}\frac{\partial{c_e^{sep}}\left ({x,t}\right )}{\partial x}}\right ], \label{eq:8}  \\ 
&\frac{\partial{c_e^{-}}\left ({\mathrm{0},t}\right )}{\partial x}=\frac{\partial{c_e^{+}}\left ({L,t}\right )}{\partial x}=0, \label{eq:9}  \\ 
&{c_e^{-}}\left ({{L^{-}},t}\right )={c_e^{sep}}\left ({{{L}^{-}},t}\right ), \label{eq:10}  \\ 
&{c_e^{sep}}\left ({{L^{-}+L^{sep}},t}\right )={c_e^{+}}\left ({L^{-}+{L^{sep}},t}\right ), \label{eq:11}  \\ 
&{D_{e,\text{eff}}^{-}}\frac{\partial{c_e^{-}}\left ({{L^{-}},t}\right )}{\partial x}={D_{e,\text{eff}}^{sep}}\frac{\partial{c_e^{sep}}\left ({L^{-},t}\right )}{\partial x}, \label{eq:12}  \\ 
&{D_{e,\text{eff}}^{sep}}\frac{\partial{c_e^{sep}}\left ({{L^{-}+L^{sep}},t}\right )}{\partial x}={D_{e,\text{eff}}^{+}}\frac{\partial{c_e^{+}}\left ({{L^{-}+L^{sep}},t}\right )}{\partial x}, \label{eq:13}
\end{align}
where ${c_e^{sep}}\left ({x,t}\right )$ represents the lithium-ion concentration in the separator, ${t_c^{0}}$ is the lithium-ion transference number, ${\varepsilon_e^{\pm}}$ and ${\varepsilon_e^{sep}}$ denote the porosities of the electrodes and the separator, respectively. Additionally,  ${D_{e,\text{eff}}}$ is the effective liquid-phase diffusion coefficient corrected by the Bruggeman relationship, and $L^{sep}$ is the separator thickness.

According to Ohm's law, the liquid-phase potential gradient can be calculated as
\begin{equation}
\frac{\partial{\phi_e^k\left ({x,t}\right )}}{\partial x}=-\frac{i_e^k\left ({x,t}\right )}{\kappa_{\text{eff}}^k}+\frac{2RT(1-{t_c^{0}})}{F} \frac{\partial \ln{c_e^k\left ({x,t}\right )}}{\partial x}, \quad k\in \{-,sep,+\}, \label{eq:14}
\end{equation}
where ${i_e\left ({x,t}\right )}$ denotes the liquid-phase current density, and $\kappa_{\text{eff}}$ represents the effective liquid-phase ionic conductivity.

At the interface between the current collectors and electrodes, current is conducted entirely by electrons, implying a zero liquid-phase current density, i.e., $i_e^{-}\left ({0,t}\right )=i_e^{+}\left ({L,t}\right )=0$.  In the separator, where no electrochemical reactions occur, the liquid-phase current density remains constant and equals the applied current density, i.e., $i_e^{sep}\left ({x,t}\right )= I\left ({t}\right )/ A$. Within the electrodes, based on charge conservation, the gradient of the liquid-phase current density equals the volumetric reaction current density, i.e., $\partial i_e^{\pm}\left ({x,t}\right )/ \partial x = {j^{\pm}}\left ({t}\right )$. Consequently, integrating Eq.~(\ref{eq:14}) along the $x$-axis yields the liquid-phase potential difference as
\begin{equation}
{\phi_e^{+}}\left ({L,t}\right )-{\phi_e^{-}}\left ({\mathrm{0,}t}\right )=-\left ({\frac{{L^{+}}}{2{\kappa_{eff}^{+}}}+\frac{{L^{sep}}}{{\kappa_{eff}^{sep}}}+\frac{{L^{-}}}{{{2\kappa}_{eff}^{-}}}}\right )\frac{I\left ({t}\right )}{{A}}+\frac{2RT\left ({1-{t_c^{0}}}\right )}{F}\ln \frac{{c_e^{+}}\left ({L,t}\right )}{{c_e^{-}}\left ({0,t}\right )}. \label{eq:15}
\end{equation}

Therefore, the terminal voltage can be expressed as
\begin{align}
V(t) &= U^{+}\left (x_{ss}^{+}(t)\right ) - U^{-}\left (x_{ss}^{-}(t)\right ) + \eta^{+}(L,t) - \eta^{-}(0,t) + \phi_e^{+}\left (L, t\right ) - \phi_e^{-}\left (0, t\right ) \nonumber \\
&= U^{+}(x_{ss}^{+}(t)) - U^{-}(x_{ss}^{-}(t)) + \frac{2RT}{F}\mathrm{arsinh}\left [\frac{j^{+}(t)}{2a_s^{+}i_0^{+}(L,t)}\right ] - \frac{2RT}{F}\mathrm{arsinh}\left [\frac{j^{-}(t)}{2a_s^{-}i_0^{-}(0,t)}\right ] \nonumber \\
&\quad - \left ( \frac{L^{+}}{2\kappa_{eff}^{+}} + \frac{L^{sep}}{\kappa_{eff}^{sep}} + \frac{L^{-}}{2\kappa_{eff}^{-}} \right )\frac{I(t)}{A} + \frac{2RT(1-t_c^{0})}{F} \ln \frac{c_e^{+}(L,t)}{c_e^{-}(0,t)}, \label{eq:16}
\end{align}
where $U^{\pm}$ is the open-circuit potential (OCP), and $x_{ss}^{\pm}(t)=c_{ss}^{\pm}(t)/c_{s,max}^{\pm}$ represents the surface stoichiometry.

\subsection{Battery degradation analysis}
During the long-term cycling of lithium-ion batteries, capacity degradation mainly stems from $\text{LAM}_{\text{PE}}$, $\text{LAM}_{\text{NE}}$, and LLI~\cite{r59,r66}. This study aims to diagnose these degradation mechanisms, thereby providing a robust physical basis for SOH estimation. This approach helps to overcome the limitations of traditional SOH assessment methods that rely on macroscopic features and lack mechanistic interpretability.

LAM refers to the loss of active material participating in electrochemical reactions. Typically, $\text{LAM}_{\text{PE}}$ originates from the dissolution of transition metal ions, and particle microcracking induced by volume expansion/contraction. These phenomena lead to electrical isolation between active particles and conductive agents or current collectors. Conversely, $\text{LAM}_{\text{NE}}$ is primarily caused by the exfoliation of graphite particles, as well as the pore blockage resulting from excessively thick SEI films. Such physicochemical damage to the microstructure directly manifests as a reduction in the effective volume fraction of active material available for lithium-ion intercalation/deintercalation. As formulated in Eq.~(\ref{eq:17}), the theoretical maximum electrode capacities $Q_{theory}^{\pm}$ are proportional to the active material volume fractions $\varepsilon_s^{\pm}$. Therefore, we select $\varepsilon_s^{\pm}$ as the key aging parameters to quantify $\text{LAM}_{\text{PE}}$ and $\text{LAM}_{\text{NE}}$.
\begin{equation}
Q_{theory}^{\pm} = A L^{\pm} \varepsilon_s^{\pm} c_{s,max}^{\pm}  F. \label{eq:17}
\end{equation}

LLI refers to the reduction in the total amount of lithium available for cycling. This phenomenon mainly stems from side reactions at the electrode/electrolyte interface. On the negative electrode, the reductive decomposition of solvent molecules leads to the growth of the SEI film, continuously consuming active lithium. Furthermore, under low-temperature or high-rate charging conditions, lithium plating may occur on the negative electrode surface, and the resulting dead lithium will no longer participate in subsequent cycles. Distinct from LAM, LLI does not alter the theoretical capacity of electrodes but shifts the stoichiometric matching between the positive and negative electrodes. Specifically, the reduction in cyclable lithium causes a relative shift in the stoichiometric operating windows, directly altering the accessible capacity. Define the surface stoichiometries at the fully discharged (SOC=0\%) and fully charged (SOC=100\%) states as $x_{0}^{\pm}$ and $x_{100}^{\pm}$, respectively. As formulated in Eq.~(\ref{eq:18}), the actual available capacity $Q_{cell}$ is determined by the stoichiometric ranges of electrodes. Therefore, we select the stoichiometric boundaries $x_{100}^{\pm}$ and $x_{0}^{\pm}$ as key parameters to characterize the LLI mechanism.
\begin{equation}
Q_{cell} = |x_{100}^{\pm} - x_{0}^{\pm}|  Q_{theory}^{\pm}. \label{eq:18}
\end{equation}

Accordingly, we construct the aging parameter vector $[\varepsilon_s^{-}, \varepsilon_s^{+}, x_{100}^{-}, x_{0}^{-}, x_{100}^{+}, x_{0}^{+}]$, which encompasses the core mechanistic parameters affecting battery capacity degradation and voltage curve distortion. Crucially, identifying these parameters bridges the gap between physicochemical degradation and macroscopic capacity fading.

\subsection{Problem formulation}
Based on the above analysis, the core task of this study is to identify the optimal aging parameters $\boldsymbol{\theta}^*$ from external measurements and subsequently estimate the SOH based on these identified parameters. 

First,  the identification of aging parameters is formulated as an inverse problem. Given the measured voltage $\boldsymbol{V}$, current $\boldsymbol{I}$, and time sequence $\boldsymbol{t}$, the optimal aging parameters $\boldsymbol{\theta}^*$ are determined by minimizing the voltage reconstruction error
\begin{equation}
\boldsymbol{\theta}^* = \mathop{\arg\min}_{\boldsymbol{\theta} \in \Omega} \left\| \boldsymbol{V} - f_{SPMe}(\boldsymbol{\theta}, \boldsymbol{I}, \boldsymbol{t}) \right\|_2^2, \label{eq:21}
\end{equation}
where $\Omega$ is the physically feasible domain of parameters, and $f_{SPMe}(\cdot)$ denotes the electrochemical model operator.

Second, as the capacity degradation is intrinsically dictated by the evolution of internal electrochemical states, the SOH is estimated as a function of the identified parameters
\begin{equation}
SOH = h(\boldsymbol{\boldsymbol{\theta}^*}). \label{eq:22}
\end{equation}
where $h(\cdot)$ denotes the mapping function leveraging $\boldsymbol{\theta}^*$ as physicochemical HIs.


However, solving this cascaded problem with high accuracy and efficiency faces significant challenges. Firstly, while integrating deep learning with electrochemical mechanisms provides a promising solution for efficient online parameter identification, it is difficult to train a network to simultaneously learn the electrochemical mechanisms and identify the aging parameters. As attempted in recent studies~\cite{r60,r55,r71}, it often causes optimization conflicts and convergence issues, which limit the accuracy of the integrated framework. Moreover, the simplified ODE constraints adopted in these studies fail to provide a reliable and accurate electrochemical basis for training the parameter identification network, further limiting the identification accuracy. Secondly, due to measurement noise and the complex coupling of degradation mechanisms, the identified parameters inevitably contain errors. Direct reliance on analytical derivations for SOH estimation can easily lead to amplified errors. 

\section{Methodology} \label{Section 3}
\begin{figure}[htbp]
    \centering
    \includegraphics[width=1\linewidth]{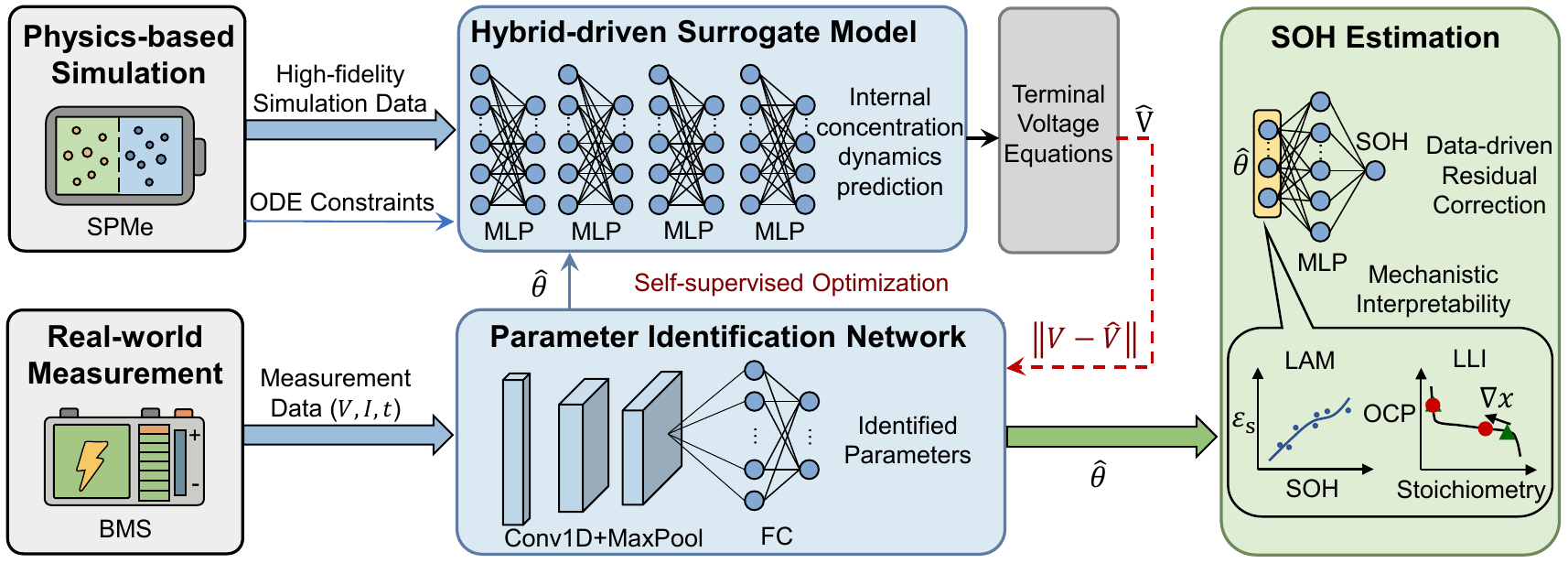}
    \caption{\label{fig:12}Schematic diagram of the proposed GPINND framework.}
\end{figure}

To tackle these bottlenecks, we propose the GPINND method that integrates deep learning with electrochemical mechanisms and adopts a sequential training strategy, as shown in Fig.~\ref{fig:12}. This approach decouples the complex heterogeneous tasks into a sequential learning pipeline, which allows each module to be specifically optimized for its unique task without compromising the overall convergence. Following this sequential training strategy, we first develop a hybrid-driven surrogate model based on high-fidelity simulation data and ODE constraints to ensure the fidelity of internal state predictions. This model serves as an accurate and differentiable physical kernel, providing the essential electrochemical basis for training the subsequent online parameter identification network. Subsequently, we construct the self-supervised parameter identification network. Utilizing the differentiable surrogate model to reconstruct terminal voltage from internal parameters, the network learns the direct mapping from external measurements to internal aging parameters by minimizing the voltage reconstruction error. Finally, using the identified parameters as robust physicochemical HIs, we train a high-precision SOH estimation network. In contrast to analytical derivations that are highly sensitive to identification errors, this data-driven mapping compensates for the identification deviations and corrects the residual errors, thereby ensuring the final estimation accuracy. Through these designs, the proposed GPINND framework ultimately achieves high-precision, low-latency aging parameter identification and SOH assessment with mechanistic interpretability.

\subsection{Hybrid-driven surrogate model}
The high computational cost of iterative parameter optimization is the primary bottleneck of online aging parameter identification. A promising solution is integrating electrochemical mechanisms with deep learning to train a neural network for direct, non-iterative parameter identification. However, recent studies that utilize simplified ODEs as constraints to construct such networks~\cite{r55,r71}, face a critical challenge. While this approach makes the electrochemical physics differentiable and suitable for gradient-based training of the parameter identification network, the physical simplification inevitably introduces systematic errors. These errors propagate and amplify through the terminal voltage equations, leading to distorted voltage reconstruction. Crucially, during loss-based parameter identification, such distortions can misguide the optimization direction, resulting in biased parameters.

To overcome this limitation, this paper proposes a hybrid-driven surrogate model combining high-fidelity simulation data with ODE constraints. This hybrid approach is specifically designed to address the systematic errors inherent in the simplified ODE constraints, thereby providing a more accurate and reliable electrochemical basis for subsequent tasks.

The terminal voltage, as indicated in Eq.~(\ref{eq:16}), is critically dependent on four key concentration dynamics $c_{ss}^{-}(t)$, $c_{ss}^{+}(t)$, $c_e^-(0,t)$ and $c_e^+(L,t)$. Considering that solid-phase and liquid-phase diffusion are governed by distinct mechanisms, using a single network to predict all concentrations simultaneously risks inducing negative transfer between tasks. Therefore, we employ four independent multilayer perceptron (MLP) subnetworks to predict the four key concentrations based on the aging parameters and operating conditions. We denote this ensemble of subnetworks as the surrogate model $\boldsymbol{\mathcal{G}}=\{\mathcal{G}_{ss}^-,\mathcal{G}_{ss}^+,\mathcal{G}_{e,0}^-,\mathcal{G}_{e,L}^+\}$. Each MLP comprises an input layer, multiple fully connected (FC) hidden layers, and an output layer, with ReLU activation functions to enhance nonlinear expressivity, as
\begin{equation}
[\hat{\boldsymbol{c}}_{ss}^{-}, \hat{\boldsymbol{c}}_{ss}^{+}, \hat{\boldsymbol{c}}_{e,0}^{-}, \hat{\boldsymbol{c}}_{e,L}^{+}] = \boldsymbol{\mathcal{G}}(\boldsymbol{\theta}, \boldsymbol{I}, \boldsymbol{t}). \label{eq:surrogate_mapping}
\end{equation}
where $\hat{\boldsymbol{c}}\in \mathbb{R}^{N \times K}$ denotes the predicted concentrations, $N$ is the number of samples, and $K$ represents the time steps.

To eliminate magnitude differences and accelerate model convergence, we employ Min-Max normalization for both input parameters and output concentrations. Specifically, for arbitrary variable $z$, the normalization is defined as
\begin{equation}
z_{norm} = \frac{z - z_{min}}{z_{max} - z_{min}}, \label{eq:23}
\end{equation}
where $[z_{min},z_{max}]$ denotes the physically feasible range of the variable $z$. 

The training of the surrogate model adopts a hybrid-driven strategy. First, extensive offline simulations are performed using the SPMe model to generate a high-fidelity dataset. This dataset covers diverse aging states throughout the full battery lifespan. Utilizing this simulation dataset for supervised training, we ensure that the model accurately captures the complex electrochemical dynamics of the full SPMe model. Furthermore, to enhance generalization, the Padé approximation method~\cite{r80} is applied to simplify the PDEs governing solid-phase and liquid-phase diffusion. The derived ODEs are subsequently incorporated as physical constraints in the loss function. Through the integration of data generation and physical constraints, we resolve the extrapolation instability of purely data-driven models and the bias of simplified physical models, enabling the surrogate model to achieve high accuracy and robust stability.

The simplified constraints governing the solid-phase concentration dynamics at the particle surface and the liquid-phase concentration dynamics at the current collectors can be respectively expressed as
\begin{align}
&\mathcal{F}_{s}^{\pm} : \quad \frac{d c_{ss}^{\pm}(t)}{d t} - \frac{3j^{\pm}(t)}{ F R_s^{\pm} a_s^{\pm}} =0 , \label{eq:24} \\
&\mathcal{F}_{e}^{\pm} : \frac{d c_{e}^{\pm}(x_{c}^{\pm},t)}{d t} - \frac{1}{\beta^{\pm}}\left[\alpha^{\pm}I(t)-\gamma^{\pm}c_{e}^{\pm}(x_{c}^{\pm},t)\right] = 0,\, \text{with } x_{c}^- = 0, \, x_{c}^+ = L, \label{eq:25} \\
&K^{\pm} = L^{-} \pm 2L^{sep} - L^{+} , \label{eq:27} \\
&\alpha^{\pm} = \frac{\pm 3(1-t_c^0)(K^{\pm} \pm 2 L^{\pm})^2}{FA} , \label{eq:32} \\
&\beta^{\pm} = L^{\pm} \varepsilon_e^{\pm}\left [3(K^{\pm})^2+10K^{\pm}L^{\pm} \pm 10 (L^{\pm})^2 \right], \label{eq:33} \\
&\gamma^{\pm} = (12 K^{\pm} + 24 L^{\pm}) D_{e,eff}^{\pm} . \label{eq:34}
\end{align}

For each subnetwork in $\mathcal{G}$, the total loss function $\mathcal{L}_{total}$ is a weighted sum of the data fitting loss $\mathcal{L}_{data}$ and the physical constraint loss $\mathcal{L}_{phy}$
\begin{equation}
\mathcal{L}_{total} = \lambda_{d} \mathcal{L}_{data} + \lambda_{p} \mathcal{L}_{phy}, \label{eq:35}
\end{equation}
where $\lambda_{d}$ and $\lambda_{p}$ are weighting coefficients.

The data loss is defined as the mean squared error (MSE) between the predicted and simulated concentrations
\begin{equation}
\mathcal{L}_{data} = \frac{1}{N K}\sum_{n=1}^{N}\sum_{k=1}^{K} \left ( \hat{c}_{n,k} - {c}_{n,k} \right )^2. \label{eq:36}
\end{equation}

The physical loss quantifies the mean-squared residuals of the governing ODEs
\begin{equation}
\mathcal{L}_{phy} = \frac{1}{N K}\sum_{n=1}^{N}\sum_{k=1}^{K} \left[ \mathcal{F}\left(\hat{{c}}_{n,k}, \frac{\partial \hat{{c}}_{n,k}}{\partial t}, I_{n,k}\right ) \right]^2, \label{eq:37}
\end{equation}
where $\mathcal{F}$ represents the physical residual operator, and the partial derivative $\frac{\partial \hat{c}}{\partial t}$ is efficiently computed via automatic differentiation. Crucially, this physical constraint provides a global regularization throughout the entire parameter space and time domain. It ensures that the predicted concentration dynamics adhere to fundamental mass conservation and diffusion mechanisms, even in parameter extrapolation scenarios.

Furthermore, to mitigate gradient dominance caused by the magnitude difference between the data loss and physical loss, we introduce a weight balancing strategy based on magnitude matching. Crucially, given the inherent systematic errors in the simplified ODEs, a lower weight prevents the physical constraints from over-penalizing the network. It allows the high-fidelity simulation data to effectively compensate for physical approximations. Based on the numerical distribution analysis during pre-training, the optimal weights are determined as $\lambda_{d}=1$ and $\lambda_{p}=0.05$. In this way, the model prioritizes learning from the accurate data to correct systematic biases, while utilizing the physical constraints primarily for regularization.

\subsection{Self-supervised parameter identification}
Leveraging the pre-trained differentiable surrogate model and terminal voltage equations, we train a self-supervised parameter identification network by minimizing the voltage reconstruction error. This approach enables the network to learn electrochemically informed aging features from measured sequences.

Specifically, we first design a parameter identification network $\mathcal{P}$ to effectively extract implicit aging features from external measurements. The network input tensor is defined as $\boldsymbol{X}=[\boldsymbol{V},\boldsymbol{I},\boldsymbol{t}]\in \mathbb{R}^{N\times K\times 3}$, where $\boldsymbol{t}$ is the normalized time sequence. The network $\mathcal{P}$ employs one-dimensional convolutional layers (1D-CNN) to capture local features from the time-series inputs. Following dimension reduction via max pooling layers, the extracted features are fed into an FC layer with the Sigmoid activation function to obtain the estimated parameters $\hat{\boldsymbol{\theta}}$, as
\begin{equation}
\hat{\boldsymbol{\theta}}  = \mathcal{P}(\boldsymbol{X}). \label{eq:41}
\end{equation}

\begin{figure}[htbp]
    \centering
    \includegraphics[width=1\linewidth]{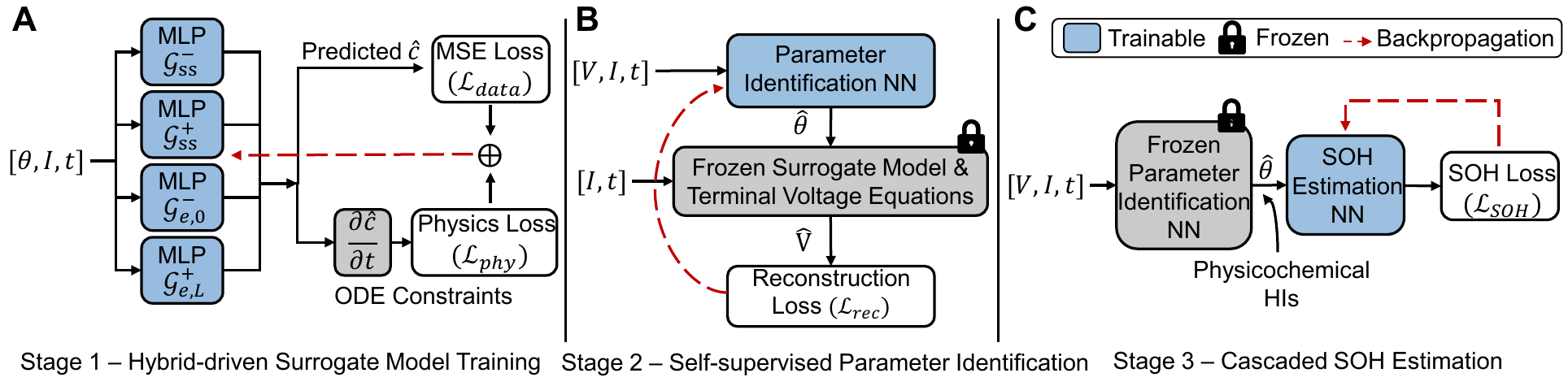}
    \caption{\label{fig:13}Schematic diagram of the sequential training strategy.}
\end{figure}

After obtaining the estimated parameters, the model proceeds to the terminal voltage calculation phase. In this stage, as shown in Fig.\ref{fig:13}, the surrogate model $\mathcal{G}$ is frozen, functioning as a differentiable physical kernel embedded within the computational graph. The estimated parameters $\hat{\boldsymbol{\theta}}$ and operating conditions (e.g., current $\boldsymbol{I}$ and time $\boldsymbol{t}$) are fed into $\mathcal{G}$ to generate the corresponding internal lithium-ion concentration dynamics. Subsequently, based on the estimated parameters and internal states, the voltage response $\hat{\boldsymbol{V}}$ is reconstructed using the terminal voltage equations of the SPMe model, denoted as the operator $\mathcal{U}(\cdot)$. The voltage reconstruction process can be expressed as
\begin{equation}
\hat{\boldsymbol{V}} = \mathcal{U} \left [ \boldsymbol{\mathcal{G}}(\hat{\boldsymbol{\theta}}, \boldsymbol{I}, \boldsymbol{t}), \hat{\boldsymbol{\theta}}, \boldsymbol{I} \right]. \label{eq:40}
\end{equation}

The MSE between the reconstructed voltage $\hat{\boldsymbol{V}}$ and measured voltage $\boldsymbol{V}$ is defined as the loss function for self-supervised training
\begin{equation}
\mathcal{L}_{rec} = \frac{1}{N K} \sum_{n=1}^{N}\sum_{k=1}^{K} \left ( \hat{V}_{n, k} - {V}_{n, k} \right )^2. \label{eq:38}
\end{equation}

During training, the gradient of the loss $\mathcal{L}_{rec}$ is backpropagated through the terminal voltage equations and frozen surrogate model to the parameter identification network, iteratively updating the network weights. This closed-loop optimization forces the parameter identification network to learn the inverse mapping from external measurements to aging parameters, ensuring the reconstructed voltage closely approximates the measured voltage. After convergence, this network enables efficient, robust, and electrochemically consistent online parameter identification. This identification offers explicit insights into the underlying degradation mechanisms and provides a solid foundation for interpretable SOH estimation.

\subsection{SOH estimation based on aging parameters}
Compared to abstract HIs, these identified aging parameters exhibit explicit, robust correlations with SOH. Therefore, we leverage the identified parameters as physicochemical HIs to improve the accuracy and interpretability of SOH estimation.

The SOH estimation network, denoted as $\mathcal{M}$, utilizes an MLP to perform non-linear regression from the identified aging parameters $\hat{\boldsymbol{\theta}}$ to SOH. To strictly enforce the physical validity, a Sigmoid activation is applied at the output layer to constrain the estimated SOH within the interval $[0,1]$.

During the training phase, we continue to adopt the sequential training strategy. As shown in Fig.\ref{fig:13}, the pre-trained parameter identification network $\mathcal{P}$ is frozen, and its output $\hat{\boldsymbol{\theta}}$ is directly fed into the SOH estimation network. Utilizing the measurement sequences $\boldsymbol{X}$ as the global input and the true SOH as the supervised label, the network minimizes the error between the estimated and true SOH. Mathematically, the SOH estimation loss is defined as
\begin{equation}
\mathcal{L}_{SOH} = \frac{1}{N} \sum_{n=1}^{N} \left [ SOH_{true, n} - \mathcal{M}[\mathcal{P}(\boldsymbol{X}_n)] \right ]^2. \label{eq:39}
\end{equation}
Based on this loss, the trainable weights of $\mathcal{M}$ are updated via backpropagation. 

This cascaded architecture achieves significant advantages by combining physical feature extraction with data-driven correction. Crucially, it not only utilizes features possessing clear electrochemical meaning as indicators, but also leverages the strong nonlinear fitting capability of NNs to optimize the SOH mapping. This design circumvents the high sensitivity to parameter deviations inherent in direct analytical calculations, as the network learns to correct and compensate for residuals. Consequently, the resulting model enables end-to-end high-precision SOH estimation from external measurements, maintaining deep physical interpretability throughout the inference process.

\subsection{Sequential training strategy}
While multi-task training is commonly used in deep learning for improving overall performance, this work adopts the sequential training strategy to ensure model convergence and physical fidelity. The proposed model-data deep fusion framework involves three heterogeneous tasks: forward internal dynamics prediction, inverse aging parameter identification, and SOH estimation. Jointly training these tasks often leads to gradient conflicts and difficulties in balancing loss weights. These convergence issues directly limit the accuracy of both parameter identification and the final SOH estimation. Critically, the simultaneous optimization can lead to parameter distortion, where the model sacrifices the physical fidelity of intermediate parameters to achieve a lower overall loss. This undermines the core purpose of integrating electrochemical mechanisms into the deep learning framework.

To address these challenges, our sequential training strategy decouples the complex non-convex optimization problem into three lower-dimensional sub-problems, as shown in Fig.\ref{fig:13}. This decoupling allows for the design of sophisticated, task-specific objective functions to achieve optimal performance on each specific task without the difficulty of balancing loss weights to ensure convergence. Furthermore, the training process follows a logical progression from learning fundamental electrochemical physics to establishing complex inverse mappings. The SOH network is subsequently optimized by leveraging the frozen identification network to ensure stable feature inputs. Therefore, this step-wise approach not only lowers optimization difficulty but also strictly preserves the physical interpretability of intermediate parameters.

For the online application, the external measurements are fed into the cascaded parameter identification and SOH estimation networks to obtain the final aging parameters and SOH, thereby realizing real-time battery health assessment with mechanistic interpretability.

\section{Experimental settings} \label{Section 4}
\subsection{Dataset}
In this work, we utilize the aging dataset of APR18650M1A LFP/graphite cells released by MIT~\cite{r72}. These batteries have a nominal capacity of 1.1 Ah and operate within a voltage range of 2.0 V to 3.6 V. The dataset contains cycling aging data from 124 batteries under diverse fast-charging strategies (single-step/multi-step fast charging with rates from 3.6C to 6C) and identical discharge conditions (4C constant current discharge to 2.0V). This diversity allows for a comprehensive evaluation of the model's applicability across various operating conditions and aging stages. The SPMe model parameters of the LFP battery are adopted from~\cite{r73}, as listed in Table~\ref{tab:1}. 

\begin{table}[htbp]
    \caption{\label{tab:1}Electrochemical parameters of the commercial LFP battery (APR18650M1A).}
    \centering
    \begin{threeparttable}
        \begin{tabular}{c c c c}
        \toprule
        Parameter & Negative & Positive & Units \\
        \midrule
        $A$ &  0.087 &  0.087   & $\text{m}^2$ \\
        $L$ & $3.5 \times 10^{-5}$ & $6 \times 10^{-5}$ & m \\
        $R_s$ & $1 \times 10^{-6}$ & $2 \times 10^{-6}$ & m \\
        $D_s$ & $3.9 \times 10^{-14}$ & $8 \times 10^{-14}$ & $\text{m}^2 \, \text{s}^{-1}$ \\
        $\varepsilon_s$ & 0.54 & 0.373 &  -- \\
        $\varepsilon_e$ & 0.40 & 0.44 & --  \\
        $x_{100}$ & 0.795 & 0.016 & --  \\
        $x_{0}$ & 0.0018 & 0.89 &  -- \\
        $c_{s,max}$ & 30555 & 22806 & $\text{mol} \,  \text{m}^{-3}$ \\
        $c_{e,0}$ & 1200 & 1200 & $\text{mol} \,  \text{m}^{-3}$ \\
        $t_{c}^0$ & 0.363 & 0.363 & --  \\
        $k_{0}$ & $3 \times 10^{-11}$ & $1.4 \times 10^{-12}$ & $\text{m}^{2.5}  \, \text{mol}^{-0.5}  \, \text{s}^{-1}$  \\
        \bottomrule
        \end{tabular}
            \begin{tablenotes}
            \item[*] Separator parameters: $L^{sep}=2 \times 10^{-5}$ m, $\varepsilon_e^{sep}=0.54$.
            \end{tablenotes}
    \end{threeparttable}
\end{table}

\begin{table}[htbp]
    \caption{\label{tab:2}The aging parameter variation ranges for the simulation dataset.}
    \centering
    \begin{tabular}{c c c c c c c}
        \toprule
        Parameter & $\varepsilon_s^-$ & $\varepsilon_s^+$ & $x_{100}^{-}$ & $x_{0}^{-}$ & $x_{100}^{+}$ & $x_{0}^{+}$ \\
        \midrule
        Range & [0.45,0.54] & [0.34,0.4] &  [0.68,0.8] & [0.0015,0.002] & [0.015,0.016] & [0.7,0.9] \\
        \bottomrule
    \end{tabular}
\end{table}

To train a high-precision surrogate model, we construct a comprehensive simulation dataset using the SPMe model. Based on the aging characteristics of LFP batteries and relevant literature~\cite{r55,r60,r71}, we defined the variation ranges for the six key aging parameters, as detailed in Table~\ref{tab:2}. This setting simulates degradation mechanisms including $\text{LAM}_{\text{NE}}$, $\text{LAM}_{\text{PE}}$, and LLI,  spanning a wide SOH range from 100\% down to 70\%.

We utilize the open-source library PyBaMM~\cite{r74} for batch simulations. To improve sampling coverage, the Latin hypercube sampling is used to generate 5,200 parameter combinations within the parameter ranges defined in Table~\ref{tab:2}. In the specific implementation, to ensure that every simulation sample strictly matches the operational voltage limits, we select $\varepsilon_s^{\pm}$, $x_{100}^-$, and $x_{0}^+$ as independent variables for Latin hypercube sampling. The remaining stoichiometric boundaries ($x_0^-$ and $x_{100}^+$) are derived based on the cut-off voltages via Eq.~(\ref{eq:16}). Subsequently, the SPMe model is executed to compute the internal lithium-ion concentration dynamics and terminal voltage for each combination under a 4C constant current discharge condition. To guarantee data quality, samples exhibiting abnormal voltage curves due to extreme parameter mismatches are filtered out. The resulting dataset includes not only typical aging trajectories but also boundary cases in the parameter space, effectively enhancing the surrogate model's robustness.

\begin{table}[htbp]
\caption{\label{tab:9}Detailed network architectures and training hyperparameters.}
\centering
\begin{tabular}{l l l}
\toprule
\textbf{Module} & \textbf{Hyperparameter} & \textbf{Configuration / Value} \\
\midrule
\multicolumn{3}{l}{\textbf{1. Surrogate Model}} \\
\cmidrule(lr){1-3} 
& Architecture & Input(8) $\xrightarrow{ReLU}$ FC(64) $\xrightarrow{ReLU}$ FC(64) $\xrightarrow{ReLU}$ FC(64) $\xrightarrow{Sigmoid}$ Output(1) \\
& Learning rate / Optimizer & $1 \times 10^{-3}$  / Adam  \\
& Batch size / Epochs & 256 / 500 \\
\midrule
\multicolumn{3}{l}{\textbf{2. Parameter Identification Network}} \\
\cmidrule(lr){1-3}
& Architecture & Input($K \times 3$) $\to$ 2$\times$(Conv1D $\to$ MaxPool) $\to$ FC(64) $\to$ Output(6) \\
& Convolutional layer &  Kernel size: 3, Filters: 16; Kernel size: 3, Filters: 32 \\
& Max pooling & Pool size: 2, Stride: 2 \\
& Activation & ReLU (Hidden), Sigmoid (Output) \\
& Learning rate / Optimizer & $1 \times 10^{-3}$  / Adam \\
& Batch size / Epochs & 64 / 500 \\
\midrule
\multicolumn{3}{l}{\textbf{3. SOH Estimation Network}} \\
\cmidrule(lr){1-3}
& Architecture & Input(6) $\xrightarrow{ReLU}$ FC(64) $\xrightarrow{ReLU}$ FC(32) $\xrightarrow{Sigmoid}$ Output(1) \\
& Learning rate / Optimizer & $1 \times 10^{-3}$  / Adam  \\
& Batch size / Epochs & 64 / 200 \\
\bottomrule
\end{tabular}
\end{table}

\begin{figure}[htbp]
    \centering
    \includegraphics[width=.9\linewidth]{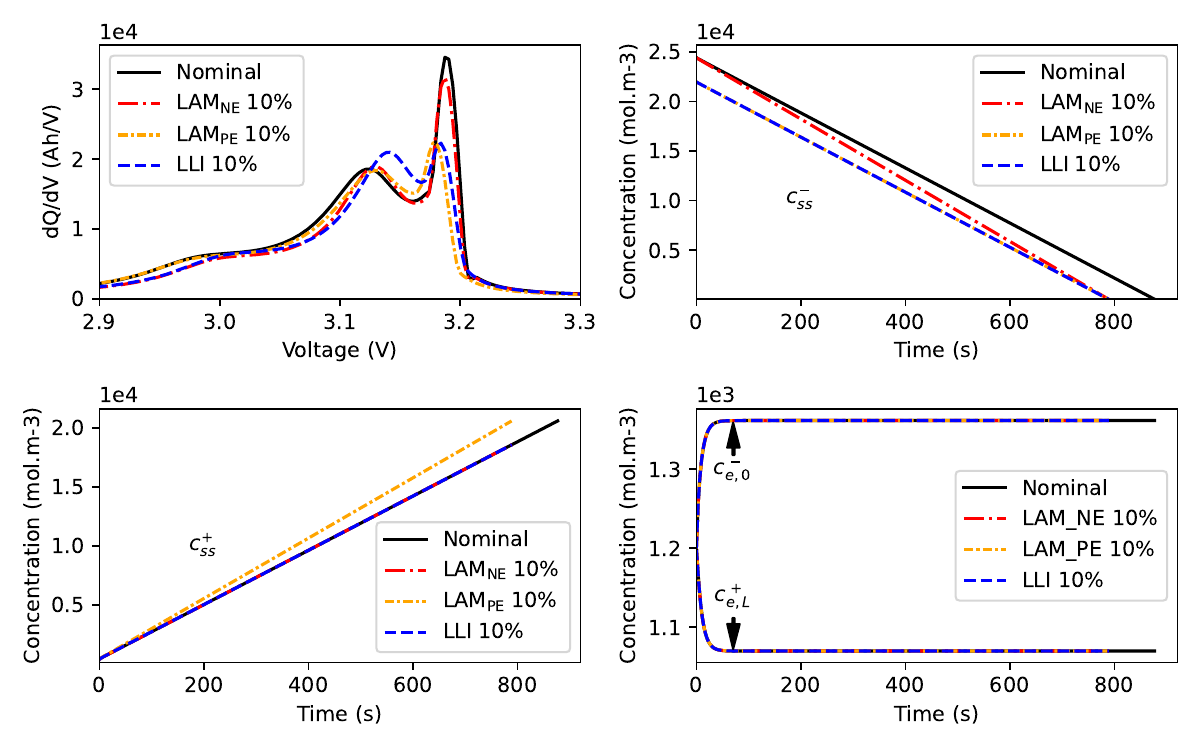}
    \vspace{-2ex}
    \caption{\label{fig:2}The impact of different degradation mechanisms on the dQ/dV curves and internal concentration dynamics.}
\end{figure}

\subsection{Baselines}
To evaluate the performance of the proposed GPINND method, we compare it with five representative SOH estimation methods, including MLP~\cite{r98}, CNN~\cite{r99}, PINN~\cite{r65}, PINN\_DeepHPM~\cite{r63}, and PIDNN~\cite{r71}.

\begin{itemize}
\item MLP: A basic data-driven baseline that leverages multiple FC layers to estimate SOH directly from input sequences.

\item CNN: A data-driven model utilizing convolutional layers to extract local temporal features from input sequences for SOH regression.

\item PINN: This method incorporates an approximated differential relationship and a monotonicity constraint as physical regularization terms to guide model training.

\item PINN\_DeepHPM: This approach utilizes a deep hidden physical model to implicitly learn battery aging dynamics, which serves as a data-discovery-based physical constraint for the SOH estimation network.

\item PIDNN: This method embeds simplified ODE constraints into NNs and employs multi-task learning to simultaneously perform internal concentration dynamics prediction, parameter identification, and SOH estimation.
\end{itemize}

\subsection{Parameter settings and experimental environment}
To evaluate the effectiveness of the proposed method and ensure reproducibility, this subsection presents the implementation details for model training and testing. For the experimental validation, both the MIT dataset and the simulation dataset are randomly split into training and test sets with a ratio of 8:2. The training set is utilized for iterative optimization of model parameters, while the test set is exclusively used for performance evaluation. Regarding model implementation, the detailed network architectures and training hyperparameters are listed in Table~\ref{tab:9}. All models are implemented using TensorFlow 2.19.0. To ensure reliability, all methods are evaluated based on the average metrics obtained from five independent trials. Furthermore, all experiments are conducted on a uniform computational platform equipped with an Intel(R) Core(TM) i9-12900K CPU @ 3.20 GHz processor and 64 GB RAM to ensure fair comparison.


\section{Results and discussion} \label{Section 5}
\subsection{Parameter sensitivity analysis}
First, we evaluate the sensitivity of key aging parameters to measured voltage, ensuring that the input tensor $\boldsymbol{X}$ contains sufficient information to capture internal battery state variations. To achieve this, we employ a local sensitivity analysis approach to assess the impact of $\text{LAM}_{\text{NE}}$, $\text{LAM}_{\text{PE}}$, and $\text{LLI}$ degradation mechanisms on the voltage response. Specifically, we reduce the corresponding active material volume fractions and stoichiometric ranges by 10\% relative to the nominal values in Table~\ref{tab:1}. Fig.~\ref{fig:2} presents the incremental capacity (dQ/dV) curves and internal concentration dynamics under different degradation mechanisms.

The dQ/dV curves presented in Fig.~\ref{fig:2} show that all three individual degradation mechanisms introduce evident deviations in the measured voltage. These deviations are fundamentally attributed to the altered internal concentration dynamics induced by specific degradation mechanisms. Analyzing the dynamics of $\boldsymbol{c}_{ss}^-$, it can be seen that under $\text{LAM}_{\text{NE}}$, the decrease in $\varepsilon_s^{-}$ directly reduces the effective active surface area. Under the constant current discharge condition, this forces the local interfacial current density to increase significantly. According to Fick's law and the Butler-Volmer equation, the increased current density intensifies the lithium-ion concentration gradient within particles, resulting in aggravated concentration polarization. Consequently, the particle surface concentration reaches the delithiation threshold prematurely, causing the terminal voltage to reach the cutoff voltage earlier. Macroscopically, this leads to a loss of discharge capacity and a distortion of the voltage profile curvature. Similarly, $\text{LAM}_{\text{PE}}$ accelerates the saturation of the positive electrode surface concentration, also leading to premature discharge termination. Conversely, LLI implies a shift of the stoichiometric operating window without significantly altering the concentration gradient because the active material volume fraction remains constant. In contrast to the significant variations in solid-phase concentration, the liquid-phase concentrations $\boldsymbol{c}_{e,0}^-$ and $\boldsymbol{c}_{e,L}^+$ exhibit relatively consistent dynamics under different degradation modes. This is primarily attributed to the sufficient electrolyte salt inventory and rapid liquid-phase diffusion. Coupled with the distinct contributions of internal concentrations to the terminal voltage, these variations of internal concentration dynamics ultimately lead to unique shape variations in the dQ/dV curves.

In summary, these distinct degradation mechanisms form discriminative measured voltage, validating the feasibility of parameter identification based on the input feature tensor $\boldsymbol{X}$.

\begin{table}[htbp]
    \caption{\label{tab:15}The RMSE of predicted concentrations for different methods.}
    \centering
    \begin{tabular}{c c c c c}
        \toprule
        Method & $\boldsymbol{c}_{ss}^-$ & $\boldsymbol{c}_{ss}^+$ & $\boldsymbol{c}_{e,0}^-$ & $\boldsymbol{c}_{e,L}^+$ \\
        \midrule
        ODE-constrained model (PIDNN) & 59.9619  &  87.1666 & 31.4801  &   81.5709 \\
        Hybrid-driven surrogate model (GPINND) &  44.3715 &  35.1552 & 0.1448  &  0.2530  \\
        \bottomrule
    \end{tabular}
\end{table}

\subsection{Concentration prediction performance}
In this subsection, we evaluate the performance of the hybrid-driven surrogate model, whose accuracy is the basis for reliable parameter identification. Table~\ref{tab:15} presents the RMSE between the predicted and simulated lithium-ion concentrations of different methods. The results indicate that the proposed hybrid-driven surrogate model achieves superior performance in predicting internal electrochemical dynamics compared to the pure ODE-constrained model. For the solid-phase concentrations, $\boldsymbol{c}_{ss}^-$ and $\boldsymbol{c}_{ss}^+$, which exhibit maximum values exceeding 20,000 $\text{mol} / \text{m}^{3}$, the surrogate model yields RMSEs of only 44.3715 and 35.1552 $\text{mol} / \text{m}^{3}$, respectively. This represents a normalized root mean square error (NRMSE) of approximately 0.002 and an accuracy improvement of at least 25\% over the ODE-constrained model, indicating that the prediction deviation is well acceptable. Notably, more significant improvements are observed in the liquid-phase concentration predictions. Since the liquid-phase concentrations exhibit relatively consistent dynamics across different aging states, the surrogate model can easily learn these patterns with high-fidelity simulation data. With magnitudes exceeding 1000 $\text{mol} / \text{m}^{3}$, the RMSEs of the hybrid surrogate model are merely 0.1448 and 0.2530 $\text{mol} / \text{m}^{3}$. This represents a substantial reduction compared to the ODE-constrained model. 

Collectively, these significant improvements indicate that the hybrid-driven training strategy effectively addresses the large errors observed in the simplified model. By leveraging the high-fidelity simulation data to correct systematic biases, the surrogate model demonstrates high prediction accuracy across arbitrary input aging parameter combinations. Crucially, this model serves as a high-precision and differentiable physical kernel, which lays a solid foundation for the self-supervised training of the parameter identification network.

\subsection{Parameter identification performance}
In this subsection, we evaluate the performance of voltage reconstruction and parameter identification using the PIDNN method as a benchmark. Table~\ref{tab:5} summarizes the voltage reconstruction RMSE of both methods on the simulation dataset and the MIT dataset (batteries $\#2$, $\#5$, $\#6$). The results demonstrate that the proposed GPINND method exhibits significant advantages under all test conditions, achieving an average voltage RMSE of only 0.0198 V, representing a 54.2\% reduction compared to PIDNN.

\begin{table}[htbp]
    \caption{\label{tab:5}The RMSE of voltage reconstruction.}
    \centering
    \begin{tabular}{c c c}
        \toprule
        Dataset & PIDNN & GPINND  \\
        \midrule
        Simulation & 0.0382 & 0.0134 \\
        MIT $\#2$ & 0.0467 & 0.0196 \\
        MIT $\#5$ & 0.0451 & 0.0205 \\
        MIT $\#6$ & 0.0426 & 0.0255 \\
        \bottomrule
    \end{tabular}
\end{table}

\begin{figure}[!ht]
    \centering
    \includegraphics[width=.9\linewidth]{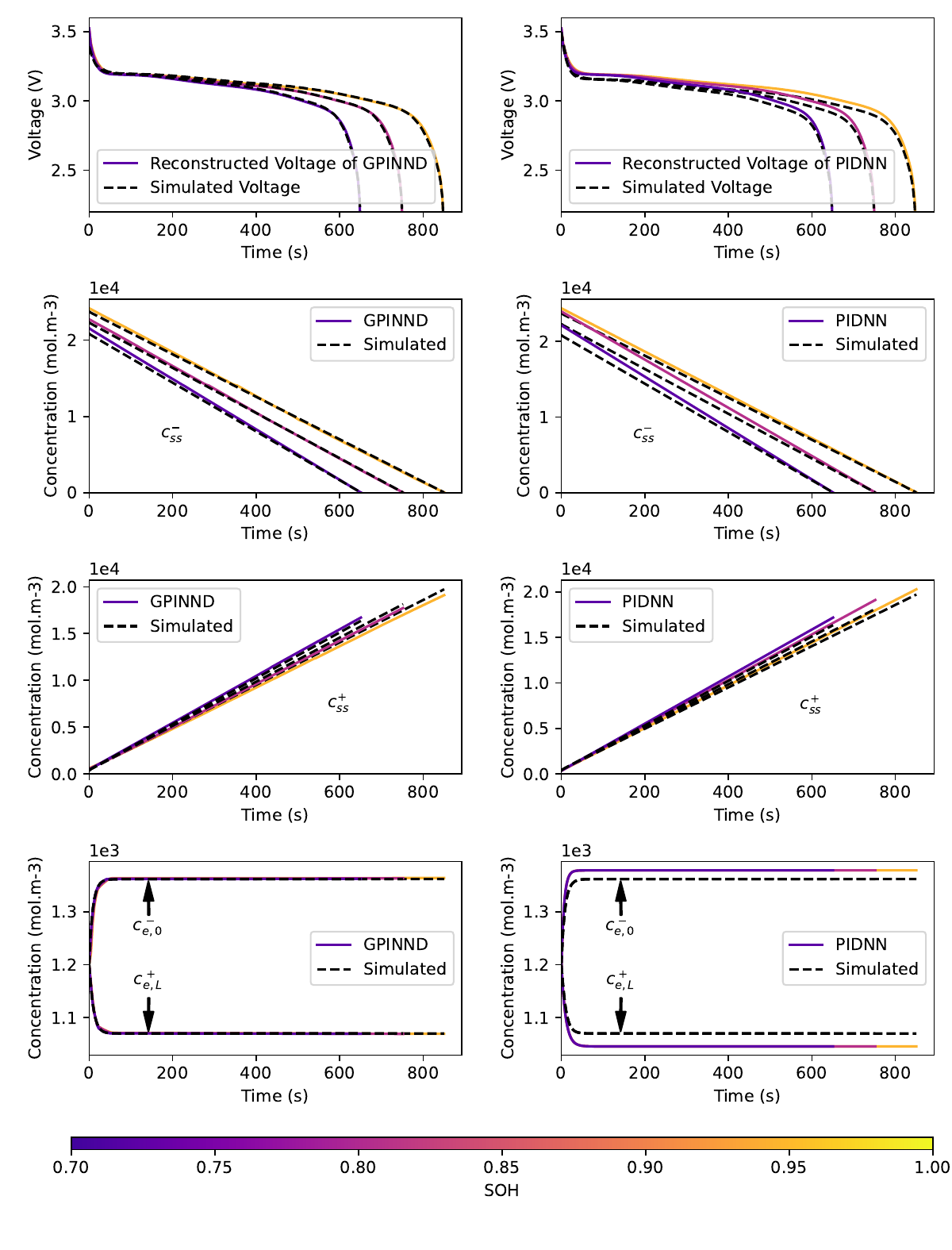}
    \vspace{-6ex}
    \caption{\label{fig:4}Comparison of reconstructed voltage and the corresponding concentration dynamics for different methods on the simulation dataset.}
\end{figure}

Fig.~\ref{fig:4} further presents the comparison of reconstructed voltage and the corresponding lithium-ion concentration dynamics of the two methods. Evidently, the proposed GPINND method demonstrates superior accuracy in both voltage reconstruction and the corresponding concentration dynamic prediction. In contrast, PIDNN exhibits larger voltage reconstruction deviations, which fundamentally reflect significant inaccuracies in both the corresponding internal concentration dynamics and the identified parameters. This performance gap stems fundamentally from the limitations of the multi-task training framework and the reliance on simplified ODE constraints. It is very complex to simultaneously optimize the aging parameters, solve the differential equation constraints dependent on these parameters, and minimize the voltage reconstruction and capacity derivation errors. This heavy multi-task burden inevitably leads to severe gradient conflicts and optimization oscillations, resulting in compromised identification accuracy. Moreover, the systematic bias introduced by simplified ODE constraints further limits the accuracy of concentration prediction and voltage reconstruction, which misguides the parameter optimization direction.


Conversely, the proposed GPINND framework overcomes these bottlenecks, owing to the precise electrochemical basis provided by the hybrid-driven surrogate model and the robust convergence enabled by the sequential training strategy. First, by utilizing high-fidelity simulation data for supervision, the surrogate model effectively mitigates the systematic bias inherent in simplified ODE constraints, ensuring accurate learning of  internal electrochemical dynamics across the entire aging parameter space. Second, the sequential training strategy isolates the parameter identification task by utilizing the frozen pre-trained surrogate model. Instead of simultaneously solving ODE constraints, this design leverages the frozen surrogate model to preserve the learned electrochemical physics while enabling efficient gradient backpropagation through the computational graph. This effectively circumvents complex gradient conflicts, enabling the electrochemically informed parameter identification from external measurements.

\begin{figure}[htbp]
    \centering
    \includegraphics[width=.9\linewidth]{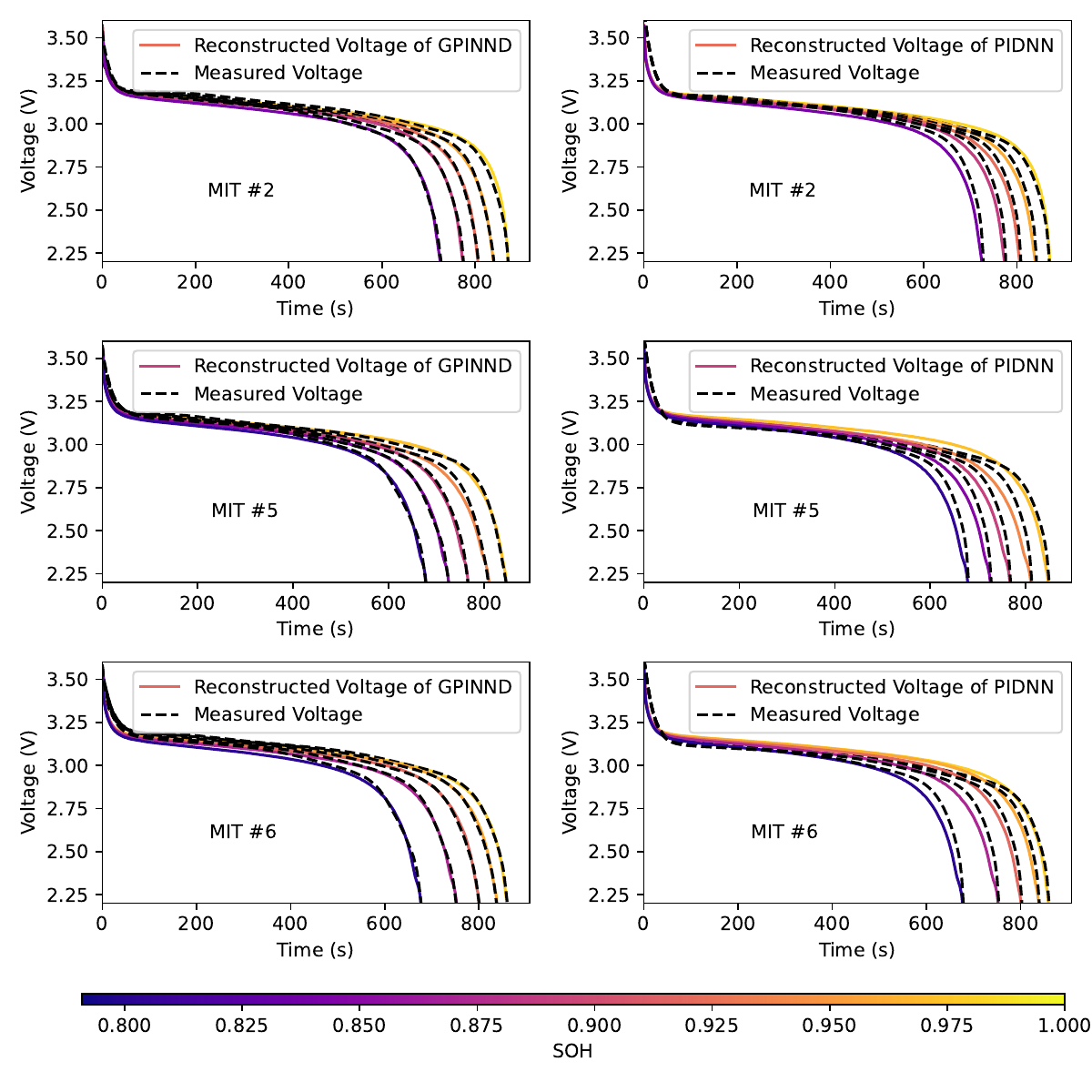}
    \vspace{-4ex}
    \caption{\label{fig:5}Comparison of the reconstructed and measured voltage for different methods on the MIT dataset.}
\end{figure}

Fig.~\ref{fig:5} illustrates the generalization performance of the proposed method on the MIT dataset. Despite the unavoidable noise in real-world scenarios, the reconstructed voltage of GPINND maintains high consistency with the measured voltage, exhibiting superior accuracy compared to PIDNN. This result strongly demonstrates the robustness of the GPINND method to environmental noise, validating its effectiveness for practical applications.



\begin{figure}[htbp]
    \centering
    \includegraphics[width=.9\linewidth]{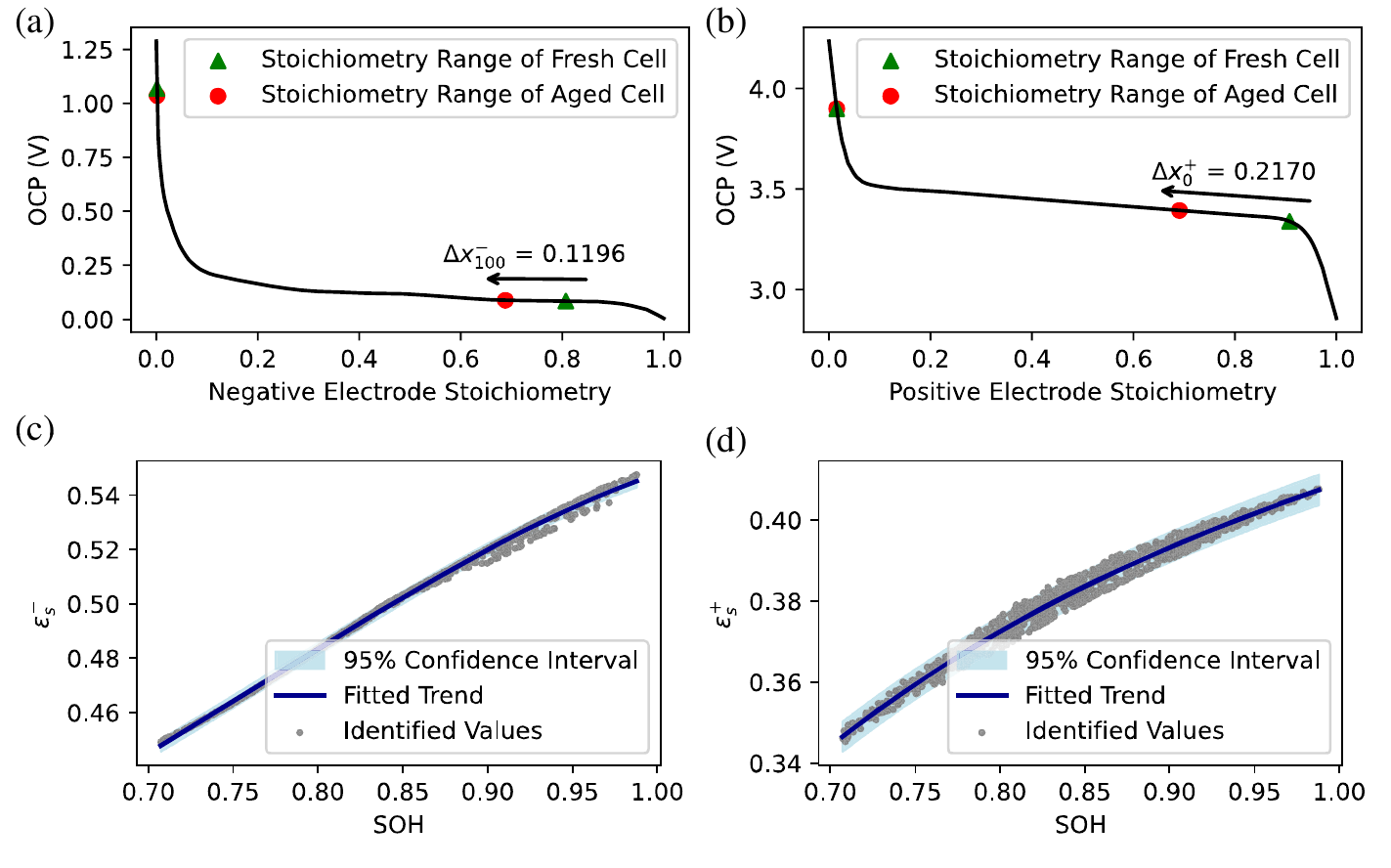}
    \caption{\label{fig:6}(a)-(b) Evolution of identified stoichiometric boundaries. (c)-(d) The relationship between the identified active material volume fractions and battery SOH.}
\end{figure}

Based on the high-precision voltage reconstruction, the model enables accurate identification of battery aging parameters. Table~\ref{tab:6} summarizes the identification RMSE of key aging parameters on the simulation dataset, where the ground truth of microscopic parameters is available. Notably, the proposed method exhibits superior identification precision compared to the PIDNN method. This improvement is attributed to the hybrid-driven surrogate model and the sequential training strategy, and it lays a solid foundation for subsequent SOH estimation based on these identified parameters. Fig.~\ref{fig:6}(a) and (b) visualize the evolution trajectory of the identified stoichiometric boundaries. As the battery ages, $x_0^+$ and $x_{100}^-$ show a significant shrinking trend, which is consistent with the mechanism of operating range shift induced by LLI. In contrast, $x_{100}^+$ and $x_0^-$ are constrained by fixed charge/discharge cutoff voltages, exhibiting only slight passive adaptive changes. Additionally, Fig.~\ref{fig:6}(c) and (d) illustrate the evolution of the identified active material volume fractions as SOH declines. It can be seen that $\varepsilon_s^{\pm}$ exhibits a strong positive correlation with SOH. These results indicate that the proposed method not only accurately reconstructs the voltage curves but also explicitly identifies the aging parameters.

\begin{table}[htbp]
    \caption{\label{tab:6}Identification RMSE of key aging parameters on the simulation dataset.}
    \centering
    \begin{tabular}{c c c c c}
        \toprule
        Method & $\varepsilon_s^-$ & $\varepsilon_s^+$ & $x_{100}^-$ & $x_{0}^{+}$ \\
        \midrule
        PIDNN & 0.0220 & 0.0267 & 0.0391 & 0.0603  \\
        GPINND & 0.0173 & 0.0265 & 0.0276 & 0.0436  \\
        \bottomrule
    \end{tabular}
\end{table}

\subsection{SOH estimation performance}
\begin{figure}[htbp]
    \centering
    \includegraphics[width=.9\linewidth]{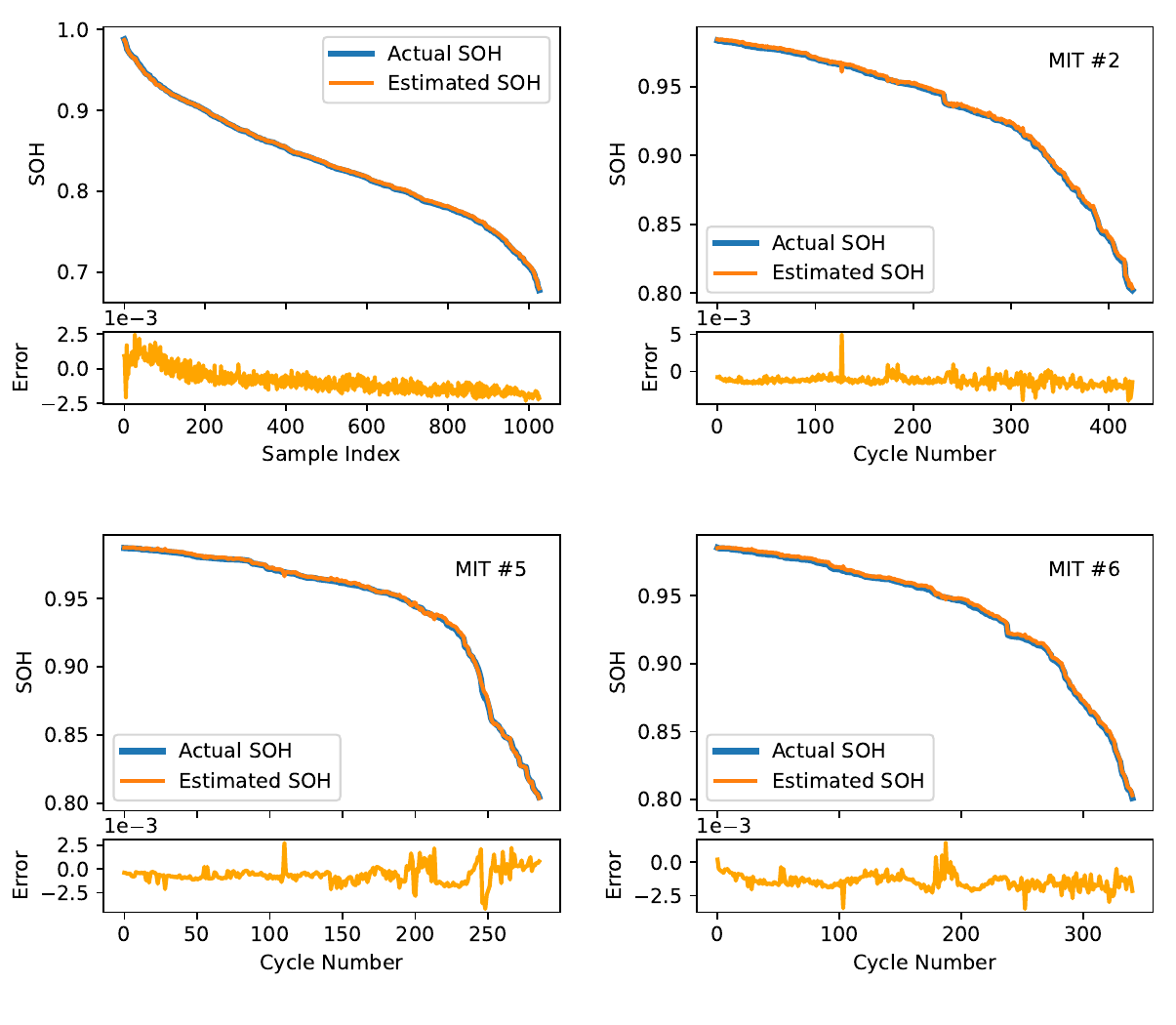}
    \vspace{-4ex}
    \caption{\label{fig:8}Estimated SOH and error of the proposed model on the simulation dataset and the MIT dataset.}
\end{figure}

Based on the identified aging parameters, we achieve SOH estimation with high accuracy and robust interpretability. Table~\ref{tab:7} summarizes the SOH estimation performance of different methods on the simulation dataset and the MIT dataset, and Fig.~\ref{fig:8} visualizes the SOH estimation results and error distribution of the proposed GPINND method.

Table~\ref{tab:7} indicates that data-driven methods, CNN and MLP, achieve an RMSE between 0.0053 and 0.0062. In contrast, the GPINND method reduces the SOH estimation RMSE by approximately 75\% compared to these methods. This significant improvement demonstrates that identifying the aging parameters through external measurements constructs a robust intermediate feature layer, thereby enhancing the accuracy and generalization of SOH estimation. Among physics-informed methods, PINN and PINN-DeepHPM show limited performance improvement, as their constraints are mostly macroscopic empirical equations or latent patterns mined from data. While PIDNN achieves an average RMSE of 0.0045 by integrating simplified electrochemical mechanisms, GPINND further improves the SOH estimation RMSE to 0.0014, representing a 68.9\% reduction compared to PIDNN. As shown in Fig.~\ref{fig:8}, the GPINND method exhibits exceptional accuracy on both the simulation dataset and the MIT dataset. The absolute errors remain below 0.005, with no significant outliers. This superior SOH estimation performance compared to PIDNN is attributed to two key advantages of the proposed framework. Firstly, the sequential training strategy allows for a cascaded parameter identification and SOH estimation pipeline. Unlike the distorted parameters that can result from multi-task training, this sequential strategy ensures a reliable, interpretable foundation for the SOH estimation. Secondly, by employing an NN for the SOH mapping, the model effectively compensates for identification deviations induced by complex degradation mechanisms and measurement noise. This data-driven correction circumvents the error amplification problem inherent in the rigid analytical derivation, significantly enhancing the final estimation accuracy. Consequently, the proposed method achieves high-precision SOH estimation while preserving robust physical interpretability.

\begin{table}[htbp]
    \caption{\label{tab:7}RMSE of estimated SOH of different methods.}
    \centering
    \begin{tabular}{c c c c c c c}
        \toprule
        Dataset & CNN & MLP & PINN & PINN-DeepHPM & PIDNN & GPINND \\
        \midrule
        Simulation & 0.0053 & 0.0058 & 0.0048 & 0.0047 & 0.0042 & 0.0014  \\
        MIT $\#2$ & 0.0056 & 0.0061 & 0.0054 & 0.0052 & 0.0044 & 0.0015 \\
        MIT $\#5$ & 0.0059 & 0.0061 & 0.0057 & 0.0057 & 0.0051 & 0.0011 \\
        MIT $\#6$ & 0.0056 & 0.0062 & 0.0053 & 0.0049 & 0.0041 & 0.0015 \\
        \bottomrule
    \end{tabular}
\end{table}

\begin{table}[htbp]
    \caption{\label{tab:10}The computational load and inference time of different methods.}
    \centering
    \begin{tabular}{c c c}
        \toprule
        Method & Computational cost (MFLOPs) & Inference Time (s) \\
        \midrule
        PIDNN & $\sim$2.1 & 0.0063 \\
        GPINND & $\sim$0.4 & 0.0012  \\
        \bottomrule
    \end{tabular}
\end{table}
\subsection{Computational Efficiency}
To validate the computational efficiency of the proposed GPINND framework, we evaluated its computational cost and inference time against the PIDNN method, as summarized in Table~\ref{tab:10}. The results clearly indicate that the GPINND method exhibits significantly higher computational efficiency. Specifically, GPINND requires only 0.4 million floating-point operations (FLOPs) and achieves an average inference time of 0.0012 s. This computational reduction compared to PIDNN is primarily attributed to the sequential training strategy, which decouples the internal dynamics prediction from the online inference process. Unlike the multi-task training framework of PIDNN, GPINND does not need to compute internal concentrations for every time step during execution. The millisecond-level inference time confirms that the proposed method is highly feasible for online parameter identification and SOH estimation integrated with electrochemical mechanisms.

\section{Conclusion} \label{Section 6}
This paper proposes a rapid battery aging parameter identification and SOH estimation method, addressing the critical challenge of reconciling computational efficiency with mechanistic interpretability. By integrating deep learning with electrochemical mechanisms and applying the sequential training strategy, the proposed method overcomes the limitations of conventional multi-task learning, offering an effective solution for online battery health management.
First, we construct a hybrid-driven surrogate model that serves as an accurate and differentiable physical kernel, which lays the foundation for training the online parameter identification network. By fusing high-fidelity simulation data with ODE constraints, the model corrects the systematic bias inherent in simplified constraints. Furthermore, through the proposed self-supervised framework, we construct an online parameter identification network informed by electrochemical mechanisms. This trained model enables direct, non-iterative identification of microscopic aging parameters from external measurements. Experimental results demonstrate that the model achieves a voltage reconstruction RMSE of 0.0198 V. Finally, by utilizing the identified parameters as physicochemical HIs and leveraging NNs for residual correction, the method achieves an SOH estimation RMSE of 0.0014, significantly outperforming existing methods. Crucially, this study demonstrates the superiority of the sequential training strategy over conventional multi-task learning for such physically coupled heterogeneous tasks. Our analysis indicates that this step-wise optimization strategy effectively mitigates gradient conflicts and facilitates the enhancement of the accuracy of each individual module.
In conclusion, this work provides a computationally efficient, physically interpretable, and highly accurate SOH estimation method for advanced BMS. Future work will extend this framework by incorporating electro-thermal coupling mechanisms and validating its adaptability across diverse battery chemistries, thereby broadening its scope for practical large-scale deployment.

\printcredits

\bibliographystyle{elsarticle-num}

\bibliography{cas-refs}

\end{document}